\documentclass[fleqn,usenatbib]{mnras}

\usepackage{newtxtext,newtxmath}

\usepackage[T1]{fontenc}

\DeclareRobustCommand{\VAN}[3]{#2}
\let\VANthebibliography\thebibliography
\def\thebibliography{\DeclareRobustCommand{\VAN}[3]{##3}\VANthebibliography}

\usepackage{graphicx}	
\usepackage{amsmath}	
\usepackage{siunitx}
\usepackage{tabularx}
\usepackage{tikz}
\usepackage{url}
\usetikzlibrary{positioning,arrows.meta,calc}

\usepackage{calrsfs}
\DeclareMathAlphabet{\pazocal}{OMS}{zplm}{m}{n}

\title[Polar Transforms for Subhalo Detection]{Polar coordinate transformations for machine learning based dark matter subhalo detection in strong gravitational lenses}

\author[J. M. Campbell et al.]{
J. M. Campbell,$^{1}$\thanks{E-mail: publications@ras.ac.uk (KTS)}
S. Dye,$^{1}$
E. Chapman$^{1}$
A. Moss$^{1}$
\\
$^{1}$School of Physics and Astronomy, University of Nottingham, University Park, Nottingham NG7 2RD, UK\\
}

\date{Accepted XXX. Received YYY; in original form ZZZ}

\pubyear{\the\year{}}

\begin{document}
\label{firstpage}
\pagerange{\pageref{firstpage}--\pageref{lastpage}}
\maketitle

\begin{abstract}
Strong gravitational lensing provides a powerful probe of dark matter, particularly on small scales where the gravitational effects of dark matter subhalos within galaxies can manifest as perturbations within the extended arcs of gravitationally lensed sources. We investigate whether transforming lensed images into polar coordinates improves the ability of convolutional neural networks to infer subhalo mass. We introduce a machine learning architecture that outputs a prediction uncertainty alongside a mass prediction to enable assessment of network reliability. Using simulated Hubble Space Telescope observations, we compare our models trained on Cartesian and polar representations under different initialisation schemes, noise levels, and subhalo concentrations ($c=60$, $c=30$). We find that polar-transformed inputs consistently yield higher subhalo detection fractions than standard Cartesian images across all tested masses. For subhalos with mass $10^9M_\odot \leq M \leq 10^{9.5}M_\odot$, the fraction of subhalos the network is able to detect increases by $\sim 15$ per cent. Pretrained networks outperform randomly initialized networks, and the polar transform consistently improves network performance in both low signal-to-noise data and for lower-concentration subhalos. The relative improvement is highest in regimes where subhalo perturbations are most difficult to detect, such as low signal-to-noise data or systems containing low concentration subhalos. These results demonstrate that presenting strong lensing images in a polar representaion provides a computationally inexpensive way of improving CNN-based subhalo detection.
\end{abstract}

\begin{keywords}
gravitaional lensing: strong -- dark matter
\end{keywords}



\section{Introduction}

The nature of dark matter remains one of the largest unsolved mysteries in the fields of astronomy and cosmology. For many years, $\Lambda$CDM has been the leading paradigm due to its ability to explain many cosmological observations \citep[For example, see reviews by][]{Primack2024, Peebles2025, Turner2026}. However, it has failed to account for observations on galactic/sub-galactic scales. One noteable example of this is the cusp-core problem, whereby dark matter simulations predict centrally cusped halo density profiles, in constrast with observations of dwarf and low surface brightness galaxies which show much shallower, cored mass distributions \citep{deBlock2010, Bullock2017}. This has motivated alternative dark matter models including warm dark matter and self-interacting dark matter \citep[See review by][]{Strigari2013}.

It is well established that, on galactic scales, dark matter forms a hierarchical structure consisting of a halo and a population of smaller subhaloes surrounding the baryonic components of the galaxy \citep{Moore1999, Springel2008, Lovell2014}. Simulations suggest that the subhalo mass function is heavily model dependent, especially at low mass ($\leq 10^9 M_\odot$) \citep{He2023}. For example, the low mass cut-off which sets the minimum mass at which subhalos can form increases with increasing mean energy of the dark matter particles. We also expect the formation of subhalos near this region to be heavily suppressed for high-energy dark matter, resulting in a severely decreased abundance \citep{Lovell2014}.  Therefore, the detection and mass measurement of these subhalos would act to constrain the subhalo mass function (SMF) and provide valuable insight into the nature of dark matter. While subhalos with sufficient mass can host luminous baryons ($\geq 10^9M_\odot$), potentially allowing us to infer subhalo mass indirectly through the luminous component, the majority are completely dark, making the task of detecting them extremely challenging, even among the Local Group. As such, the only viable probe for subhalo detection is through gravitational influence on luminous matter and light.

In the local environment, subhalo searches have been extensively conducted through a variety of techniques including stellar weak lensing \citep{VanTilburg2018, Mondino2020}, and analysis of extended cold stellar streams \citep{Erkal2015, Erkal2016, Bonaca2025}. While these techniques have had success, they cannot be extended to higher redshifts due to the need for high resolution observations. The only gravitational technique that can be applied to high redshift systems is galaxy-galaxy scale gravitational lensing. This phenomenon occurs when two galaxies are approximately aligned in the line-of-sight direction. Light from the farthest (source) galaxy is deflected by the closer (lens) galaxy, forming a distorted and magnified image of the source in the form of lensed arcs. The resulting image of the source is governed by the morphology of the source galaxy and the mass distribution of the lens galaxy, including dark matter components. The presence of any subhalo with a projected position in the lens plane on or near the Einstein ring will therefore create small perturbations within the image. The magnitude of these perturbations depends on the mass of the subhalo.

One way to exploit gravitational lensing for substructure detection is through Bayesian modelling. Using this technique, the lens is first modelled with only a smooth mass distribution. Substructure is then added in order to improve the likelihood of the fit. This technique has been used successfully to detect and measure the mass of subhalos in several instances. The first was made by \citet{Vegetti2010} who detected a subhalo with mass $M_{200} \sim 10^{10}M_\odot$ in the galaxy SLACS J0946+1006. Additionally, \citet{Vegetti2012} reported the discovery of a subhalo with mass $M_{200} \sim 10^{9}M_\odot$ in the galaxy JVAS B1938+666. While more recent re-analysis of this system has revealed the object to be a foreground field halo rather than a galaxy subhalo, it still speaks to the efficacy of the approach. More recently, \citet{Amvrosiadis2026} reported a $M_{200} \sim 10^{9}M_\odot$ subhalo in the galaxy PJ011646. However, the Bayesian method does have several drawbacks. Firstly, the modelling process is very slow and computationally expensive, growing exponentially more complex as substructures are added to the model. To properly constrain the mass function, subhalo searches need to be conducted across a large number of systems, which this technique is not optimised for. Additionally, as the perturbations created by subhalos are extremely small, the detections are extremely sensitive, and heavily rely on accurate modelling of both the smooth and substructure mass components \citep{oRiordan2024}. This makes Bayesian modelling a very manual process, and further adds to the overall run time. Furthermore, the masses obtained from Bayesian modelling are dependent on the assumed subhalo mass profile; inferred masses can be biased by over a magnitude \citep{Minor2017}. 

Due to the shortcomings of the Bayesian method, there has been considerable work done in recent years in applying machine learning to this problem. Such approaches generally fall into one of two categories. The first aims to directly detect individual subhalos by searching for perturbations in Einstein rings, for example using a binary classification CNN \citep{DiazRivero2020} or image segmentation methods \citep{Ostdiek2022, Tsang2024}. In all cases, single subhalo detectors are consistently able to identify high mass subhalos positioned proximate to the Einstein ring. This method has the advantage of being able to measure both the position and mass of subhalos directly, and can detect multiple subhalos in one image, but generally fails to identify lower mass subhalos ($\leq 10^9 M_\odot$) required to differentiate between model-dependent mass functions.

The alternative approach to single substructure detection is characterising the effects of a population of subhalos on lensing images. This can be advantageous as it doesn't require detection of individual subhalos because the collective effect of a population of substructures is typically larger than a single subhalo. This method is therefore not subject to the same data quality constraints as individual subhalo detection. \citet{Brehmer2019} utilised a neural ratio estimator to obtain the abundance and slope of the SMF used to generate simulated images. \citet{Filipp2025} used the same method to determine how such models cope with inference of lensing images with parameters outside the parameter space of the training images. \citet{Varma2020} successfully used a classifier to determine the low mass cutoff of a series of simulated data. As well as pure subhalo detection, population level inference can be used for other substructure tasks. \citet{Alexander2020} used classification to differentiate between different types of dark matter substructure in lensing images.

There have been several instances where polar coordinate transformations have been used to increase the effectiveness of gravitational lensing tasks. \citet{Joseph2014} employed a PCA-based algorithm that utilises polar transforms to construct an automated strong lens finder. A subsequent approach for the same task used a machine learning model combined with similar coordinate transformations \citep{Hartley2017}. They argue that the Gabor filters used in their network can better exploit tangential features of lenses when the image is polar transformed. While no direct comparisons are presented using identical methods without polar transformations, the results nevertheless suggest that such transforms may be broadly useful across multiple lensing tasks.

In this work, we investigate the effect of polar transformations on a standard machine-learning-based subhalo detection approach. We perform a direct comparison of network performance between CNN-based architectures trained on standard strong lens images and an identical network trained on polar-transformed strong lens images. We conduct this comparison using an adapted version of a ConvNeXt-tiny, as well as looking at the effects of using pretrained weights.

This paper is organised as follows. In Section \ref{sec:methods} we outline our methods, including the generation of simulated data, model architectures and network training. In Section \ref{sec:results} we show the results of our comparison and in Section \ref{sec:discussion} we consider reasons for the observed behaviour as well as implications for future work. Finally, we present our conclusions in Section \ref{sec:conclusions}.

\section{Methods}\label{sec:methods}

\subsection{Data Generation}

Training CNN-like architectures to detect subhalos within strong lensing images most naturally lends itself to supervised learning. Therefore, we require a labelled training dataset. 

For this work, we simulate strong lensing images with a single smooth mass component including shear and multipoles, a single smooth source, and a single subhalo. We generate mock strong lensing images using the \textsc{lenstronomy}\footnote{\url{https://github.com/lenstronomy}} python package \citep{Birrer2018}. To generate realistic lensing observations, we utilise the \textsc{lenstronomy-simulation api} which allows for image generation according to real instrument configurations. For this work, we choose to simulate observations of the Hubble Space Telescope Wide Field Camera 3 Infrared (WFC3 IR) F160W filter. This accounts for pixel scale and point-spread function (PSF). The signal-to-noise ratio (SNR) in the image is determined by the source magnitude as well as the exposure time (or equivalently the number of HST orbits). The distribution of source magnitudes used in this study is shown in table \ref{tab:lensing_params} \citep{Ferrami2024}. 

We conduct our analysis across two separate groups of networks: one trained and evaluated on high-SNR data and one trained and evaluated on low-SNR data. The high-SNR data is generated using 50 HST orbits and the low-SNR data is generated using a single HST orbit. All other aspects of the datasets are identical. We keep the redshift of the lens and source fixed at $z_{lens}=0.2$ and $z_{source}=0.6$. Each image contains 64$\times$64 pixels. The pixel scale of the WFC3 IR instrument is $0.13\arcsec$ per pixel, however we assume a drizzeled pixel scale of $0.08\arcsec$ per pixel. 

\begin{table}
    \centering
    \begin{tabular*}{\linewidth}{l @{\extracolsep{\fill}} r}
         \hline
         Parameter & Distribution \\
         \hline
         \hline
         \multicolumn{2}{c}{Smooth Lens} \\
         \hline
         Einstein radius, $\theta_E$ & $U[0.8, 1.5]\arcsec$ \\
         Ellipticity, $q$ & $U[0.4, 1]$ \\
         Orientation, $\phi$ & $U[-\pi, \pi]$\\
         Position, $(x, y)$ & $U[-0.25, 0.25]\arcsec$ \\
         \hline
         \multicolumn{2}{c}{Shear} \\
         \hline
         $\gamma_1$ & $U[-0.2, 0.2]$ \\
         $\gamma_2$ & $U[-0.2, 0.2]$ \\
         \hline
         \multicolumn{2}{c}{Multipoles} \\
         \hline
         $m$ & $3, 4$ \\
         $a_m$ & $U[0, 1]$ \\
         \hline
         \multicolumn{2}{c}{Source Light} \\
         \hline
         Magnitude & $U[17, 21]$ \\
         Radius, $R_{\text{Sérsic}}$ & $U[0.1, 0.8]$ \\
         Sérsic Index, $n_{\text{Sérsic}}$ & $U[0.7, 1]$\\
         Ellipticity, $q$ & $U[0.33, 1]$\\
         Orientation, $\phi$ & $U[-\pi, \pi]$\\
         Position, $(x, y)$ & $U[-0.1, 0.1]$\\
         \hline
         \multicolumn{2}{c}{Substructure} \\
         \hline
         Mass, M & $logU[10^{7}, 10^{11}]M_\odot$ \\
         Concentration, c & $60$ \\
         Truncation Scale $\tau$ & $5$\\
         \hline
    \end{tabular*}
    \caption{Lens parameter distributions of the simulated observations used to train the networks.}
    \label{tab:lensing_params}
\end{table}

Parameter distributions used for this work are largely based on \citet{Ostdiek2022} however we choose to simplify some aspects to allow for easier training and comparison of our networks. We model the smooth component of the lensing mass as a Singular Isothermal Ellipse (SIE) \citep{Kormann1994}, parameterised by the Einstein radius ($\theta_E$), ellipticity ($q, \phi$) and position ($x, y$). We model the source light distribution as a Sérsic ellipse, parameterised by magnitude, radius ($R_{\text{Sérsic}}$), Sérsic index ($n_{\text{Sérsic}}$), ellipticity ($q, \phi$) and position ($x, y$).

In addition, we add shear and both $m=3$ and $m=4$ multipoles to the lens model to simulate more realistic lens configurations. Shear approximates tidal gravitational fields typically found in lens environments due to external sources of mass. Neglecting this would mean simulating lenses in a completely empty environment. Multipoles act to deviate the smooth lens component from complete ellipticity. Not only does this produce more realistic lenses, it also accounts for the effect that limited flexibility in the mass component of the lens can lead to false subhalo detections \citep{Nightinghal2024, Lange2025}. For future work where networks will be applied to real data, it is important that the network avoids false positives due to lacking angular complexity in the training data. We therefore ensure in this work that any improvements made to performance through polar transformations apply to data consisting of multipoles. In this work, the amplitudes of both the $m=3$ and $m=4$ multipoles are drawn from the uniform distribution $U[0,1]$.

The subhalos are modelled using a truncated NFW (tNFW) profile \citep{Navarro1996, Baltz2009}:

\begin{equation}
    \rho (r) = \frac{M_0}{4\pi r \left(r+r_s\right)^2} \left(\frac{r_t^2}{r^2+r_t^2}\right)
\end{equation}

\noindent where $M_0$ is proportional to the total mass, $r_s$ is the scale radius and $r_t$ is the truncation radius. To obtain the total mass of the profile, $\rho(r)$ is integrated from $r=0$ to $r=r_t$ We use a truncation scale $\tau = r_t/r_s = 5$. The scale radius and concentration parameter relate via $r_s=R_{200}/c$, where $R_{200}$ is the radius at which the mass density is 200 times the critical mass density. In \textsc{lenstronomy}, this profile is parameterised by mass, $M$ and concentration, $c$, and the other parameters are calculated from this. For this work, we first choose a concentration parameter of $c=60$. We note that this is higher than values generally predicted by $\Lambda$CDM, however aligns with recent observations of high concentration substructure \citep{Minor2021, Sengul2022}. We also note \citet{Tsang2024} found machine learning networks performed adequately on data with $c=60$ subhalos, but failed completely when applied to $c=15$ subhalos. Considering this, we conduct further testing on $c=30$ subhalos (see Section \ref{sec_low_conc}). The chosen parameter values/distributions are shown in Table \ref{tab:lensing_params}. For a subhalo to create a large enough perturbation to be detected, it must be placed such that its projected position is within close proximity to the Einstein ring. We therefore place the subhalo randomly within the region where the pixel brightness is within $0.5\times$ the peak brightness of the lensed image.

We frame this problem as a regression task, therefore the target labels take the form of a continuous distribution of masses. We set each label as $\hat{y}=\log_{10} \left( M/M_\odot \right)$. One limitation of this approach is that it does not naturally accommodate images with no subhalo in the training set. This is because the target is only defined when a subhalo is present. Adding a series of training samples where $\hat{y}=0$ would create a severe discontinuity within the training set and could adversely affect network training \citep{Kowatsch2024}. Our training dataset therefore only contains images with exactly one subhalo, as the aim of this work is to assess only whether polar transforms improve network performance. Nevertheless, as we show in section \ref{sec:results}, our models readily indicate if a substructure has not been reliably identified.

\subsection{Network Architecture}

\begin{figure*}
    \centering
    \begin{tikzpicture}[
    node distance=0.45cm,
    box/.style={
        draw,
        rounded corners,
        align=center,
        minimum height=1.3cm,
        minimum width=2.2cm,
        font=\small
    },
    bigbox/.style={
        draw,
        rounded corners,
        align=center,
        minimum height=1.8cm,
        minimum width=4.8cm,
        font=\small
    },
    outbox/.style={
        draw,
        rounded corners,
        align=center,
        minimum height=1.1cm,
        minimum width=1.8cm,
        font=\small
    },
    arrow/.style={->, thick}
]

\node[box] (input) {
Input\\
$3 \times 64 \times 64$
};

\node[bigbox, right=of input] (encoder) {
\textbf{ConvNeXt-Tiny Encoder}\\[0.1cm]
Stage 1: $96$, $3$ blocks \hspace{0.3cm}
Stage 2: $192$, $3$ blocks\\[0.05cm]
Stage 3: $384$, $9$ blocks \hspace{0.3cm}
Stage 4: $768$, $3$ blocks
};

\node[box, right=of encoder] (gap) {
Global Avg.\\
Pooling
};

\node[box, right=of gap] (head) {
MLP Head\\
$768 \rightarrow 512$
};

\node[outbox, above right=0.45cm and 0.7cm of head] (mu) {
$\mu$\\
$512 \rightarrow 1$
};

\node[outbox, below right=0.45cm and 0.7cm of head] (logvar) {
$\log \sigma^2$\\
$512 \rightarrow 1$
};

\draw[arrow] (input) -- (encoder);
\draw[arrow] (encoder) -- (gap);
\draw[arrow] (gap) -- (head);

\draw[arrow] (head.east) -- ++(0.35,0) |- (mu.west);
\draw[arrow] (head.east) -- ++(0.35,0) |- (logvar.west);

\end{tikzpicture}
    \caption{The architecture of the network used in this work, consisting of a ConvNeXt feature encoder and custom linear heads. The overall architecture consists of a series of convolutional stages and downsampling stages, as well as an initial stem that converts each non-overlapping 4×4 region of the input image into a learned feature vector, producing a downsampled feature map.}
    \label{fig:convnext_feature_extractor}
\end{figure*}
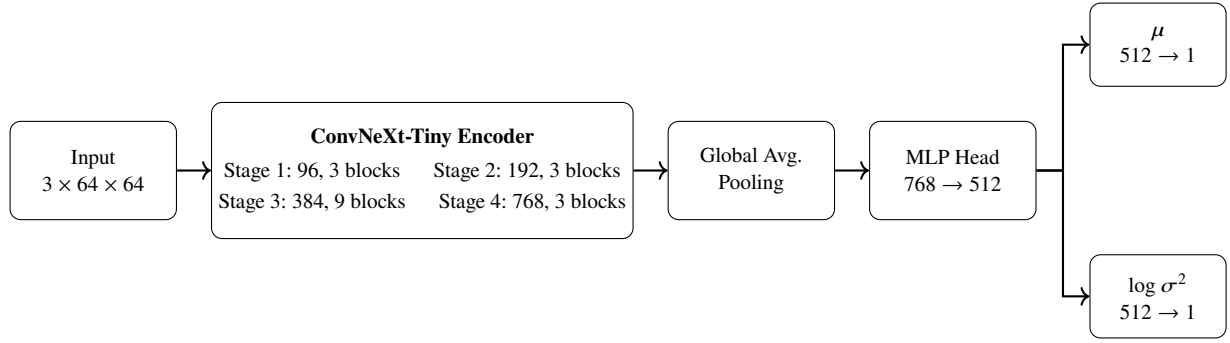

For image vision tasks, CNN-based architectures have been the standard for over a decade. For this work we use a custom architecture derived largely from the recently developed ConvNeXt-Tiny \citep{Liu2022}. CNN architectures can be separated into two main parts; the feature encoder and the linear head(s). The feature encoder utilises convolutional layers to extract features from the images in the form of abstract feature maps. The linear head then aims to map these features onto output vectors, which vary depending on the task.

For our network, we use the feature extractor of the ConvNeXt-Tiny architecture. The input image is first processed by the "patchify" stem, where a 4$\times$4 convolution with stride 4 downsamples the image by a factor of 4, and projects them onto a 96 channel feature space. The remainder of the feature encoder consists of several convolutional stages, comprising a series of ConvNeXt blocks. Each ConvNeXt block contains a depthwise 7$\times$7 convolution, followed by Layer Normalization, two pointwise linear layers with GELU activation, and a residual connection. Between each layer is a downsampling layer that increases the number of feature channels, thus providing a more complete feature representation. The channel dimensions of the four blocks are 96, 192, 384, 768 respectively. Figure \ref{fig:convnext_feature_extractor} shows a diagram of the ConvNeXt-Tiny architecture used in this work.

The final encoder feature map is reduced using global average pooling and passed to a multilayer perceptron (MLP) regression head. In regression tasks, the network gives an output value for every input vector. In the case of this work, it outputs a prediction of the subhalo mass within each image. For subhalos in lensing images however, we expect that low mass perturbations will not provide a large enough perturbation to be detectable by the network. In this case, the network must still produce a mass estimate, making it difficult to distinguish between predictions that are supported by information in the image and those that are correct by chance. We therefore design an MLP linear head that is capable of outputting both a mass prediction and a prediction (aleatoric) uncertainty. Aleatoric uncertainty represents the irreducible uncertainty inherent in the observations, arising from measurement noise, observational limitations, or intrinsic ambiguity in the data. In our framework, the uncertainty predicted directly by the network is interpreted as an estimate of this uncertainty. The MLP head consists of a fully connected layer mapping 768 nodes to 512 nodes, followed by a ReLU activation and dropout (p=0.1). Two independent linear output layers are then used to predict the mean $\mu$ and log-variance log($\sigma^2$) of the target distribution, enabling heteroscedastic uncertainty estimation through a Gaussian negative log-likelihood loss.

Typical regression networks use a loss function such as Mean Squared Error (MSE) which is a function of just the network prediction. The linear head would therefore output only the network prediction. We use an alternative loss function, Negative Log-Likelihood (NLL) which naturally includes a variance term. Rather than simply training the model to predict a single value for subhalo mass, NLL loss trains the model to predict the parameters of a probability distribution for the mass. Assuming a Gaussian distribution of mean $\mu$ and variance $\sigma^2$, the likelihood of observing a true value, $\hat{y}$, under this distribution for a single observation is

\begin{equation}\label{eq:likelihood}
    p(\hat{y}|x) = \frac{1}{\sqrt{2\pi\sigma^2}}\exp{\left(-\frac{(\hat{y}-\mu)^2}{2\sigma^2}\right)}.
\end{equation}

\noindent For $N$ images, the total likelihood function is the product of all individual likelihood functions:

\begin{equation}
    p(\hat{y}_1,...,\hat{y}_N|x_1,...,x_N) = \prod_{i=1}^{N}p(\hat{y}_i|x_i)
\end{equation}

\noindent The final NLL loss is therefore

\begin{equation}\label{eq:nll_loss}
    \begin{split}
        \pazocal{L} = -\log{\prod_{i=1}^Np(\hat{y}_i|x_i)} 
         = \frac{1}{N}\sum_{i=1}^{N} \left[ \frac{(\hat{y}_i-\mu_i)^2}{2\sigma_i^2} + \log{\sigma_i} \right].
    \end{split}
\end{equation}

\noindent The $\frac{1}{2}\log{2\pi}$ term that arises from normalisation of eq.~(\ref{eq:likelihood}) is omitted from eq.~(\ref{eq:nll_loss}) as this does not change the gradient of the loss. The loss function for each image now consists of two terms: an error term and an uncertainty term. The error term takes the form of a standard MSE loss function, scaled by a variance term. The mean determines the accuracy of the prediction, and the variance controls how strongly a large prediction error is penalised. The uncertainty term acts to regularise the loss function, preventing the network  predicting excessive uncertainties in all cases to minimise the loss.

To accommodate this loss function, we design a linear head which outputs both $\mu$ and $\log{\sigma^2}$. It takes the 768 channel feature map that is outputted from the feature extractor and converts it to a 768 dimensional vector via adaptive average pooling. This vector is then processed by a fully connected layer which projects the 768 dimensional vector into 512 dimensions. The reduced dimensionality acts as a compact bottleneck, lowering the parameter count of the prediction head while maintaining sufficient capacity for the regression task. We then apply a ReLU activation and dropout with a probability of $p=0.1$ for regularisation. The resulting vector then branches into two separate linear layers; one predicts the mean of the target distribution, and the other predicts the log-variance. This means the network outputs both variables required for the loss function as separate trainable parameters.

\subsection{Polar Transformation} \label{sec:polar_transform}

Gravitationally lensed images are naturally structured around a central mass, leading to features that are approximately radial or azimuthal. To better exploit this structure, the data can be transformed from Cartesian to polar coordinates, where each pixel is represented by its radius $r$ and angle $\theta$ relative to a defined centre.

\begin{figure*}
    \centering
    \includegraphics[width=\linewidth]{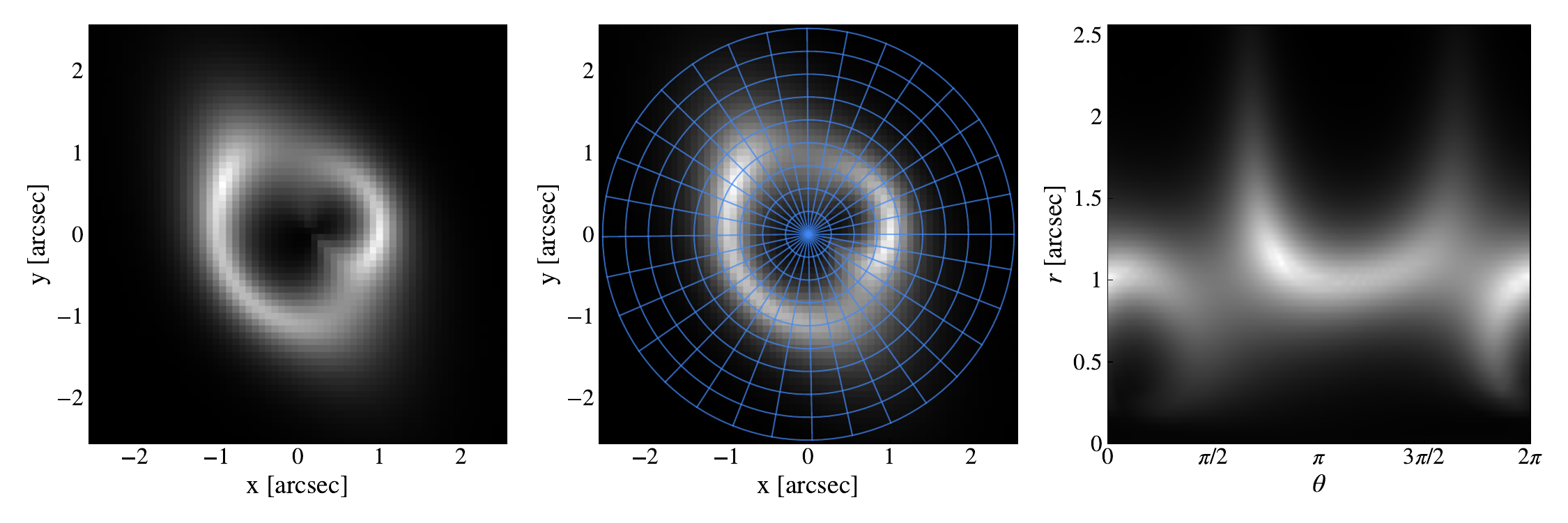}
    \caption{An example of applying the polar transformation to an image. The left panel shows the original simulated observation from the dataset. The middle panel shows an example of the resampling grid overlayed on the original image. For visualization purposes, this grid has a lower resolution than what was used during the training process. The right panel shows the resulting polar-transformed image.}
    \label{fig:polar_transform_demo}
\end{figure*}

The initial Cartesian images are taken from the original dataset, thus guaranteeing an unbiased comparison between networks. To convert an image from Cartesian to polar representation, we overlay a polar resampling grid and, for each ($r$, $\theta$) position, obtain a pixel value via the \textsc{pytorch} implementation of bilinear interpolation. We choose the sampling grid to have 64 uniformly distributed values of $r$ from $0$ to $2.56\arcsec$, and 64 uniformly distributed $\theta$ values from $0$ to $2\pi$. We choose these bins in order to ensure the resulting image has the same dimensions as the Cartesian image, ensuring a fair comparison between networks. Due to the parameter distributions used to create each observation, we choose to always centre the grid at $(0,0)$ as every lens is approximately central. An example of this process is shown in Figure \ref{fig:polar_transform_demo}.

\subsection{Network Training}

In addition to testing the effect of polar transforms, we want to determine the importance of pretraining in this problem. In machine learning, pretraining involves training a network on a large image dataset prior to application to the target problem. The pretrained network develops feature extraction layers that are sensitive to common image structures, allowing subsequent training on the scientific dataset to converge more rapidly and often achieve improved performance compared to training from random initial conditions.

To assess the effect of polar transformations and the impact of pretraining, four models were trained per dataset corresponding to all combinations of coordinate representation (Cartesian or polar) and initialisation (random initialisation or pretrained). Henceforth, we abbreviate the four network types as CRI, PRI, CPT and PPT. For the pretrained networks, we initialise the ConvNeXt feature encoder using \textsc{IMAGENET\_1K} weights \citep{Deng2009}. These pretrained weights are not optimized for astronomical data, however they encode generic low-level visual features such as edges, gradients, and simple textures. Such features are largely domain-agnostic and can facilitate more efficient optimization when fine-tuned for a specific task. In order to apply these pretrained weights, the network expects a 3-channel input format. Since our data is single-channel, we duplicate the image across all three channels to match this requirement without altering the underlying information. We also do this for the networks with random initializations in order to ensure a fair comparison.

For each dataset, we generate a total of 500,000 images. Of these, 80 per cent are designated for training, 10 per cent for validation, and the final 10 per cent for testing. We use the AdamW optimizer, with a weight decay of $\num{1e-3}$. The initial learning rate is selected independently for each network to ensure stable and effective optimization; we observe that networks trained with excessive initial learning rates fail to learn anything significant. For pretrained networks, we used an initial learning rate of $\num{2e-4}$. For randomly initialized networks, we found that an initial learning rate of $\num{1e-5}$ allowed for successful learning in a reasonable time frame. Although the initial learning rate varied, each network followed the same learning rate schedule throughout the training process: if the validation loss hasn't improved for 15 epochs, the learning rate is dropped by a factor of 10, with a minimum of $\num{1e-7}$. This scheduling strategy allows the optimiser to take relatively large steps during the early stages of training, enabling rapid progress towards a good region of the parameter space. As training progresses and the loss approaches a minimum, smaller learning rates help stabilise optimisation and allow the network to make finer adjustments to the model parameters. Training stops when the validation loss has not improved for 50 epochs.

It is common for deep neural networks such as those used in this work to overfit when trained on limited datasets. To mitigate this issue, we apply data augmentation throughout training. Augmenting the data increases the diversity of the input images, making it much harder for the model to learn specific examples seen during training, instead learning appropriate features that correspond to the labels. For each training image, we apply a random rotation between $-\pi$ and $\pi$, using bilinear interpolation. Each image also has a 50 per cent chance of being flipped horizontally as well as a 50 per cent chance of being flipped vertically on top of the initial rotation. These augmentations are valid as gravitational lenses are invariant under rotation and reflection; rotating or mirroring a lensing image does not alter the underlying physical lensing configuration.

\section{Results}\label{sec:results}

\begin{figure}
    \centering
    \includegraphics[width=\linewidth]{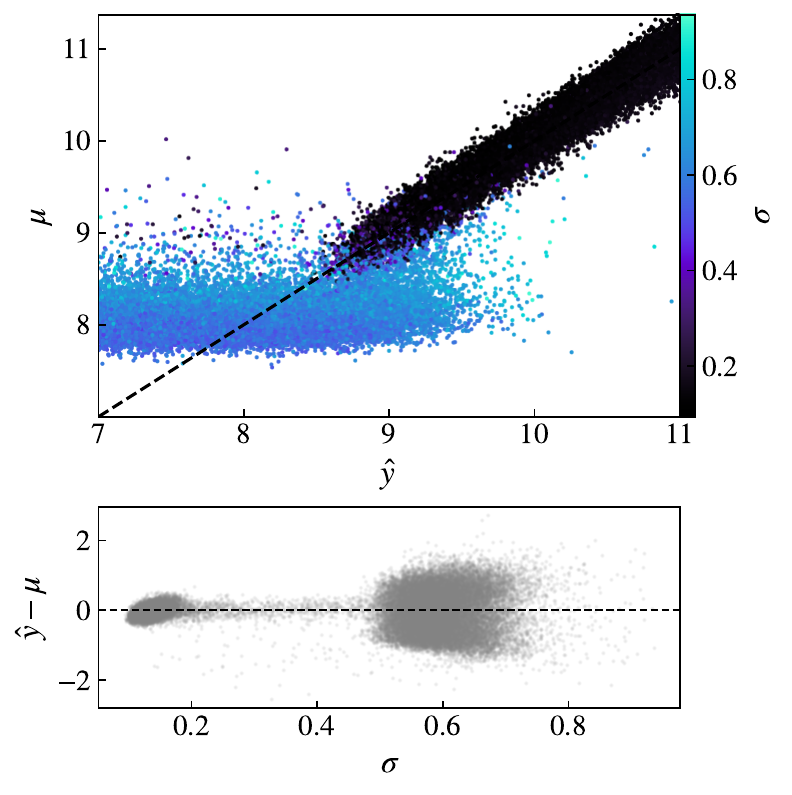}
    \caption{The output of the pretrained network trained on polar-transformed images when used to evaluate a testing dataset. The top panel shows the predicted subhalo mass against the true subhalo mass. The colour represents the uncertainty given by the network for each prediction. The bottom panel shows the prediction error ($\hat{y}-\mu$) against the predicted uncertainty.}
    \label{fig:network_outputs}
\end{figure}

To characterise the performance of each network, we run all images in the testing dataset through the network and obtain a mass prediction and uncertainty for each. We only test the network on the same type of data with which it was trained (i.e. coordinate representation, noise level, and as described later, lensing parameters). Figure \ref{fig:network_outputs} shows an example of the distribution of predictions and uncertainties for the testing dataset, though all networks show similar behaviour. It is clear that the network predictions can be divided into two distinct populations according to the prediction error and uncertainty; a diagonal branch where the predictions accurately trace the true values at high masses, and a horizontal branch where the network fails but guesses an approximate median value to minimize the MSE term in the loss function. 

\begin{figure}
    \centering
    \includegraphics[width=\linewidth]{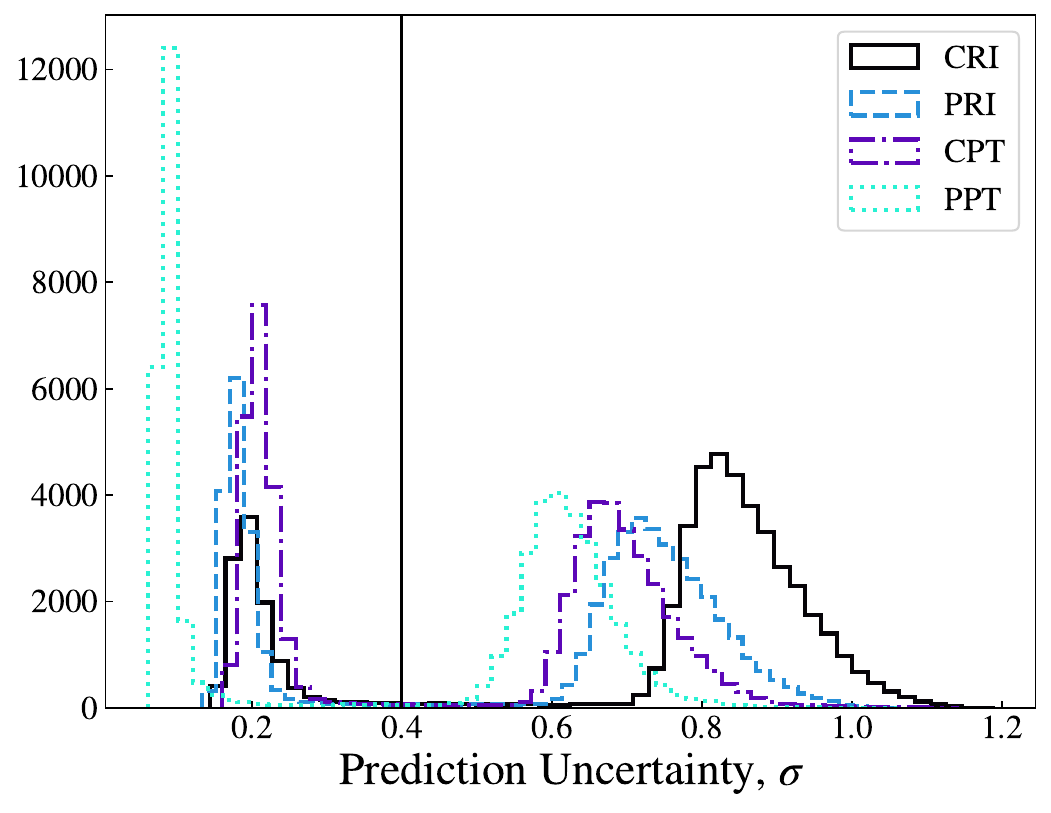}
    \caption{The distribution of prediction uncertainties across the testing dataset for each of the four networks. The black line indicates the chosen threshold for determining the confidence of a network's prediction. The legend entries denote Cartesian - random initialization, polar - random initialization, Cartesian - pretrained, polar - pretrained respectively. The uncertainties can be well approximated as two separate populations, with a threshold of $\sigma=0.4$ dividing the two.}
    \label{fig:uncertainty_hist}
\end{figure}

To distinguish between accurate and inaccurate predictions, we set thresholds on both the prediction error and uncertainty. For the prediction error, we set the threshold at $|\hat{y}-\mu|<0.5\;\text{dex}$. This gives a relatively conservative condition that the predicted mass must lie within one order of magnitude of the true value. The significant part of determining a successful detection comes from the uncertainty. For the uncertainty threshold ($\sigma$), we consider the distribution of uncertainties across the networks. These distributions are shown in Figure \ref{fig:uncertainty_hist}, which shows two clearly distinct populations for each network. The emergence of a bimodal uncertainty distribution indicates that the network is not assigning uncertainties continuously or arbitrarily, but is instead classifying predictions into confident and uncertain regimes. This behaviour is consistent with a well-calibrated model that recognizes when the input does or does not constrain the target variable. We choose a threshold of $\sigma < 0.4$ as this splits the two populations for all networks. If a prediction is below both of these thresholds, we label it as detected, otherwise we label it as not detected. While not explicitly shown, we have confirmed that these thresholds are also valid for subsequent networks trained in this work.

\subsection{Detection Fraction}

To determine the networks' ability to detect substructure of different masses, we bin the predictions by true mass with bin width $0.5\;\text{dex}$. For each mass bin, the detection fraction is defined as the ratio of the number of detected subhalos in that bin, $k$, to the total number of subhalos in that bin, $N$.

\begin{figure}
    \centering
    \includegraphics[width=\linewidth]{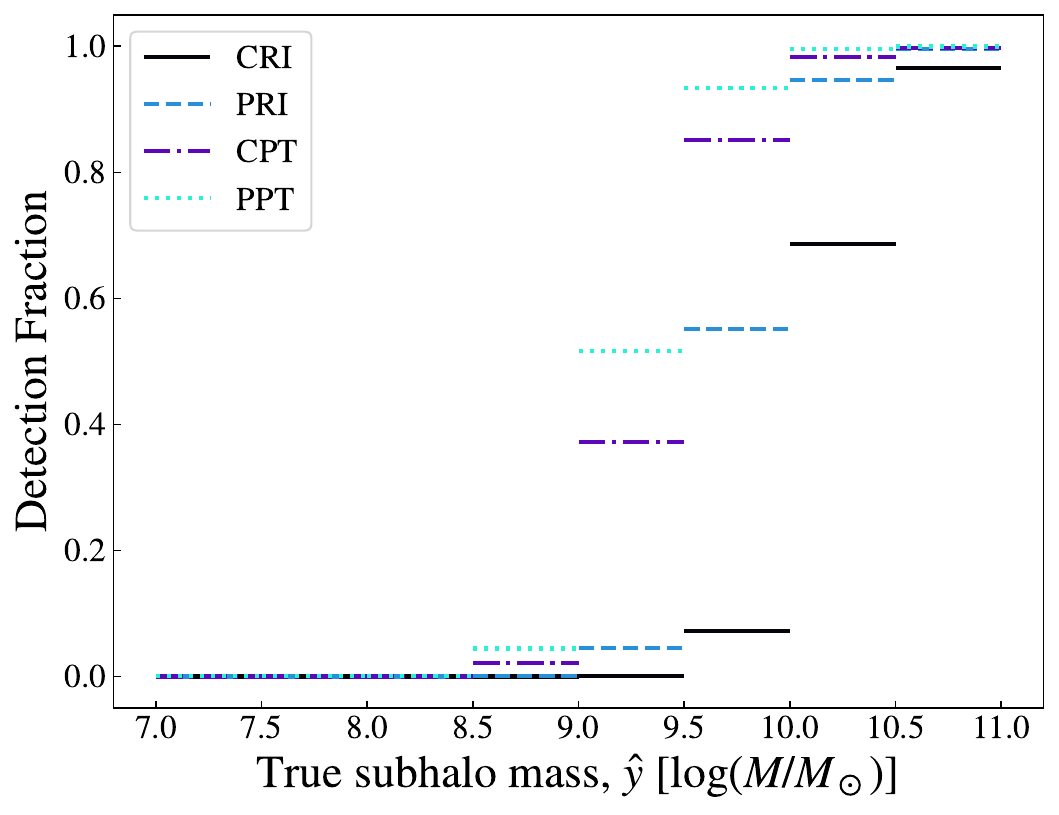}
    \caption{The detection fraction of subhalos as a function of true mass for the four different networks. The legend is as described in Figure \ref{fig:uncertainty_hist}. For both the pretrained and randomly initialized cases, the networks using polar-transformed data outperform the Cartesian networks.}
    \label{fig:model_comparison}
\end{figure}

Figure \ref{fig:model_comparison} shows the detection fraction as a function of subhalo mass for the four networks. Each network shows a similar overall behaviour. The detection fraction is essentially zero in the low mass regime as any perturbation signal produced by the subhalo either falls below the noise level or its spatial extent is far below the pixel scale of the image. As the mass increases, so too does the detection fraction, increasing rapidly then levelling off. At the very highest masses ($M\sim10^{11}M_\odot$) the detection fraction is approximately unity. It is clear that network performance is significantly affected by both pretraining and coordinate representation. Pretraining significantly enhances performance in the intermediate mass regime, particularly for the polar representation, which shows an earlier and steeper increase in detection fraction compared to its non-pretrained counterpart. In contrast, the Cartesian models lag behind, requiring higher masses to reach a comparable detection fraction. In the most extreme case, changing the coordinate representation from Cartesian to polar increases the detection fraction by $\sim 50$\,per cent for the non-pretrained case, and $\sim 15$\,per cent for the pretrained case.

Given that polar-transformed networks outperform Cartesian networks across both pretrained and randomly initialized models, and that pretrained models consistently surpass their randomly initialized counterparts, subsequent comparisons are conducted using only pretrained networks.

\subsection{Low Signal-To-Noise Data}

We now look at how images with significantly lower signal-to-noise affect the performance gain from polar transforms. For this, we generate a new full dataset using a single HST orbit. All other parameter distributions are kept as listed in Table \ref{tab:lensing_params}. We follow an identical procedure of both training and analysis as done for the 50 orbit data, using the same thresholds to determine whether a detection is valid. We use a lower learning rate than for the previous networks, as we find that lowering the initial learning rate to $\num{1e-5}$ allows for much more stable training and better performing networks in this regime.

\begin{figure}
    \centering
    \includegraphics[width=\linewidth]{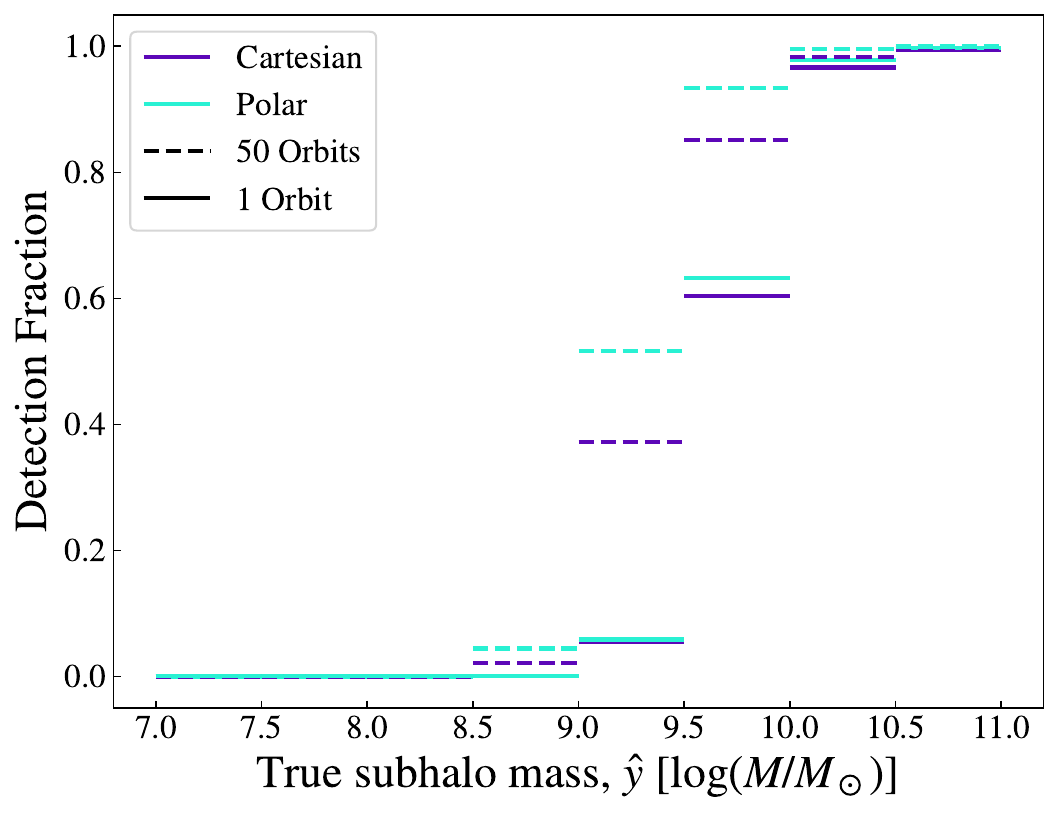}
    \caption{The detection fraction for the pretrained polar and Cartesian  networks. The solid lines represent the networks trained and evaluated on data generated with a single HST orbit, and the dashed lines represent the networks trained and evaluated on data generated with 50 HST orbits. Despite lower overall performance with respect to high signal-to-noise data, the polar transform retains its advantage.}
    \label{fig:single_orbit_comparison}
\end{figure}

Figure \ref{fig:single_orbit_comparison} shows the detection fraction (DF) for the two pretrained networks as a function of true subhalo mass as well as comparing them to the high SNR counterparts. The figure clearly shows that, while the polar network does have superior performance over the Cartesian network, the gain is substantially diminished in comparison with the high signal-to-noise networks. The polar networks retain between $1-5$\,per cent higher detection fraction in all bins where $0 \lesssim {\rm DF} \lesssim 1$ as opposed to the $\sim15$\,per cent gain in the low SNR regime. While the improvement introduced by the polar transform is smaller, this is expected, as the polar transform improves the efficiency with which lensing features are represented, but does not increase the information content of the data. In the low SNR regime, detection becomes information-limited as perturbation signals drop below the noise level, and the advantage of an improved representation diminishes.

\begin{figure*}
    \centering
    \includegraphics[width=\linewidth]{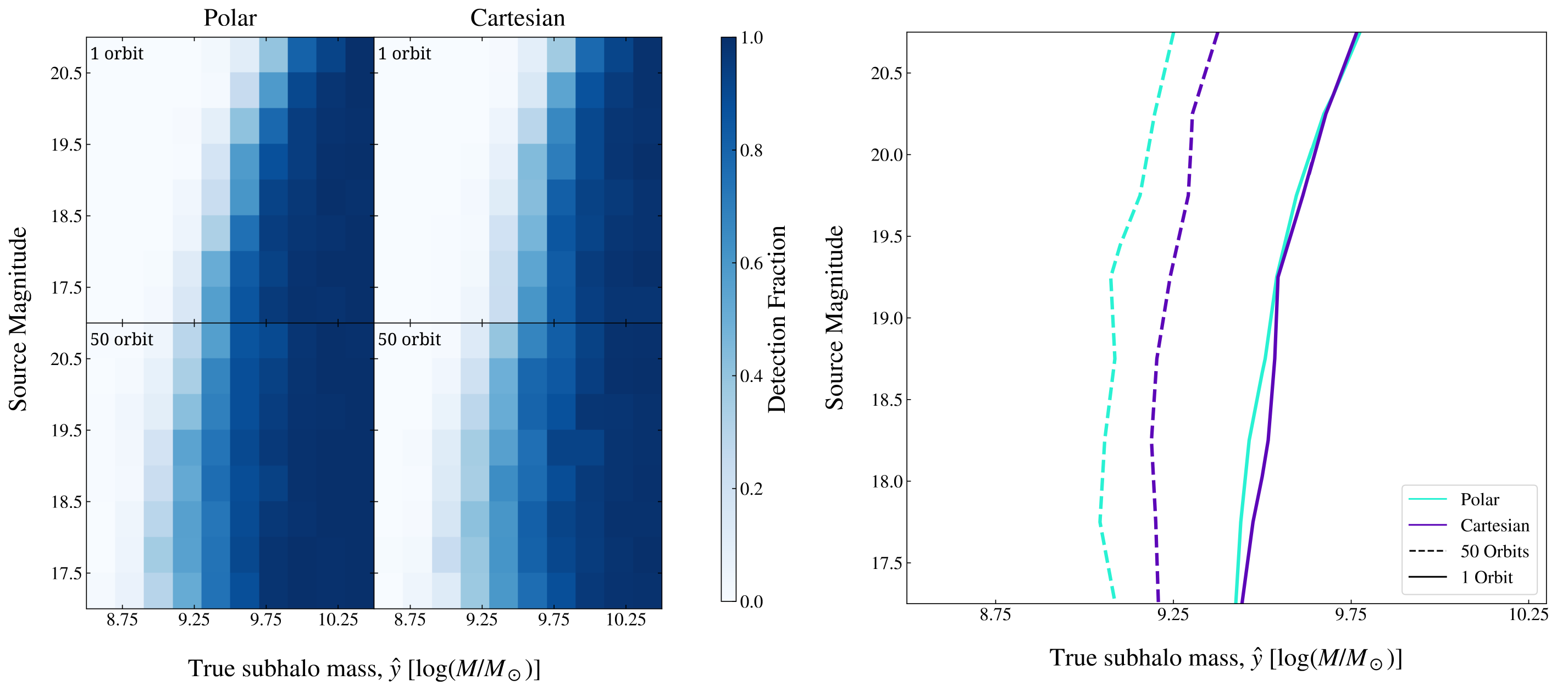}
    \caption{Left: Heatmaps of the detection fraction of the PPT and CPT networks trained and evaluated on single orbit data (top) vs 50 orbit data (bottom). Right: Contours of the 50\% detection fraction for each of the four networks. Only the region of parameter space where the detection fraction varies is shown. As source magnitude decreases, and the signal-to-noise increases, the mass at which the network obtains a 50 per cent detection fraction decreases. This decrease is sharper for the networks using polar-transformed data.}
    \label{fig:mag_mass_heatmaps}
\end{figure*}

Given that the noise level of the data is determined by total flux in the image, which is correlated with the source magnitude, we also examine how the detection fraction varies with source magnitude in the two noise regimes. The left panel of Figure \ref{fig:mag_mass_heatmaps} shows a heat map of the detection fraction as a function of both source magnitude and true subhalo mass for the polar and Cartesian networks in both noise cases. The right panel shows the 50\,per cent detection fraction contour for each network. 

As has already been shown, the overall detection fraction for the polar networks is superior in both regimes, however loses most of its advantage in the low SNR regime. For high SNR data, the improvement from the polar transform is approximately constant across source magnitudes. For the low SNR data, however, the only improvement that can be seen occurs for the lowest source magnitudes, where the SNR is the highest. This suggests that as the SNR decreases, whatever is driving the improvement through the polar transform becomes less pertinent. Since these are 50\,per cent detection contours, this cannot be explained by the noise completely washing out the perturbation signal. It therefore suggests that the features or lensing configurations that are allowing for detections at this level do not benefit from a change in coordinate representation.

\subsection{Low Concentration Subhalos}
\label{sec_low_conc}

So far, we have used a subhalo concentration of $c=60$ when making simulated images. While recent potential subhalo discoveries through Bayesian modelling suggest subhalos with comparable or even far greater concentrations can exist \citep{Quinn2021}, it is generally accepted that expected concentrations of subhalos in the $\Lambda$CDM paradigm are lower ($c=15-30$) \citep{Ludlow2016}. Given this, we also check that benefits provided by polar-transformed data are still present when analysing data containing lower concentration subhalos. To do this, we generate a new dataset with the same parameter distributions as in Table \ref{tab:lensing_params}, but with a subhalo concentration $c=30$. We then train and evaluate two new networks, one for Cartesian data and one for polar data, and compare as before.

\begin{figure}
    \centering
    \includegraphics[width=\linewidth]{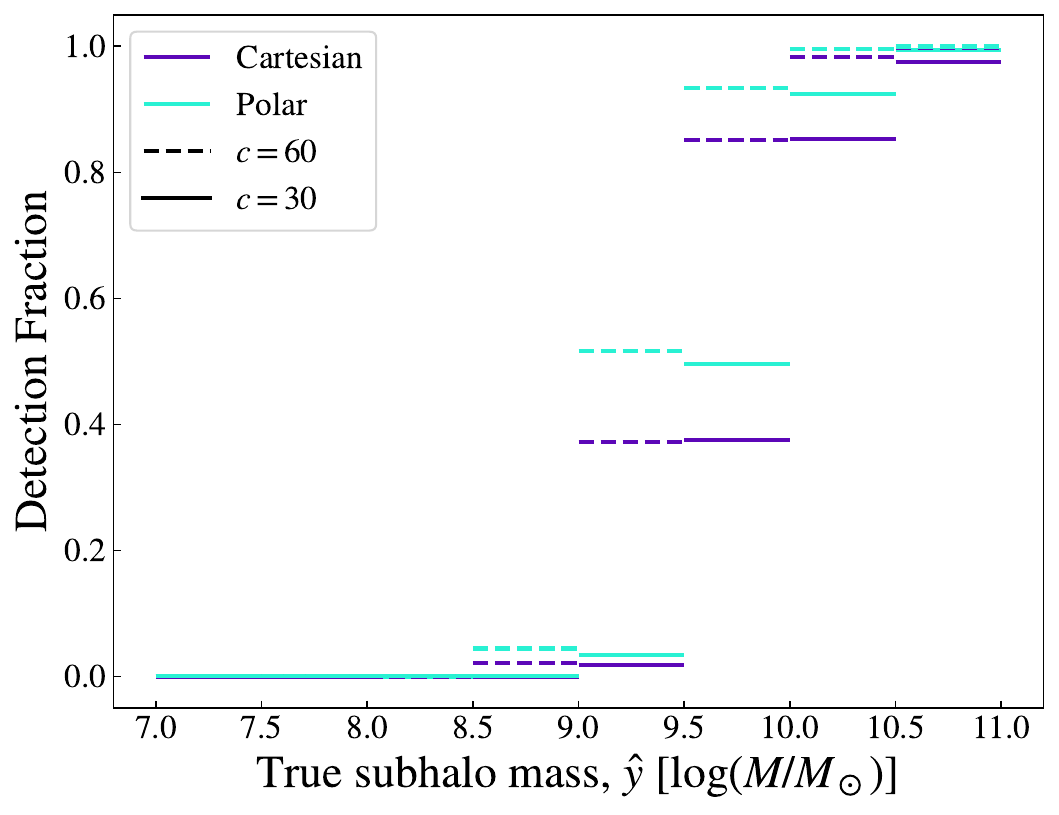}
    \caption{The detection fraction for the pretrained polar and Cartesian networks. The solid lines represent the networks trained and evaluated on data generated using $c=30$ subhalos, and the dashed lines represent the networks trained and evaluated on data generated using $c=60$ subhalos. Despite lower overall performance with respect to high subhalo concentration data, the polar transform retains its advantage.}
    \label{fig:network_comparison_concentration}
\end{figure}

Figure \ref{fig:network_comparison_concentration} shows the detection fraction for both the Cartesian and polar networks for data generated with $c=30$ subhalos. The high concentration counterparts are also shown for comparison.

The trends observed for the $c=60$ subhalo population remain relatively unchanged for the lower-concentration $c=30$ sample. In both cases, the polar-transformed network achieves higher detection fractions at fixed mass and reaches the transition to efficient detection at lower masses than the Cartesian network. Reducing the concentration shifts the overall detection threshold to higher masses, consistent with the weaker perturbations generated by less centrally concentrated subhalos, but does not remove the advantage of the polar representation. This indicates that the improvement from the polar transform is robust across different subhalo structural properties.

\section{Discussion}\label{sec:discussion}

We have shown that a convolutional-based network trained to measure the mass of subhalos within galaxy scale strong lensing images performs better when the data are first transformed to a polar representation. While the initial testing was done in idealised conditions (high signal-to-noise, high concentration subhalos), we have also shown that the benefits provided by the polar transform persist when extended to low signal-to-noise data and when detecting lower concentration subhalos. The fact that the increase in performance from the polar transform is consistent across all three regimes suggests that it is improving the accessibility of information required to determine subhalo mass.

\subsection{Why do polar transforms improve performance?}

The consistent improvement observed when using polar-transformed inputs suggests that this alternate representation presents information to the network in a way more conducive to finding subhalos. In general, Einstein rings exhibit a strong degree of axial symmetry. In Cartesian space, they appear as highly curved structures in all parts of the image. This means that convolutional kernels, which operate on rectangular grids and are themselves rectangular, have to learn rotationally varying structure. Therefore, identically physical structures that appear at continuously changing orientations appear different to the network. On the other hand, polar transforming the inputs partially linearises lens morphology causing similar morphological features and subhalo perturbations to appear more consistent. This allows the convolutional layers to better learn features that indicate subhalo perturbations.

Additionally, the polar transform rearranges the arcs into a more linearly coherent structure. Since convolutions are performed by sliding local kernels across the image, the linearised lensing features are more naturally aligned with the receptive fields of standard convolutional layers. This allows local kernels to sample neighbouring regions of the Einstein ring more coherently, potentially improving the extraction of weak subhalo perturbations.

\subsection{Quantifying the improvement across regimes}

To assess the relative improvement introduced by the polar transform, we examine the statistic $(DF_{polar}-DF_{cart})/(DF_{polar}+DF_{cart})$, which takes positive values when the polar transform yields an increased detection fraction compared to the Cartesian grid and negative values when the standard Cartesian grid performs better. 

\begin{figure}
    \centering
    \includegraphics[width=\linewidth]{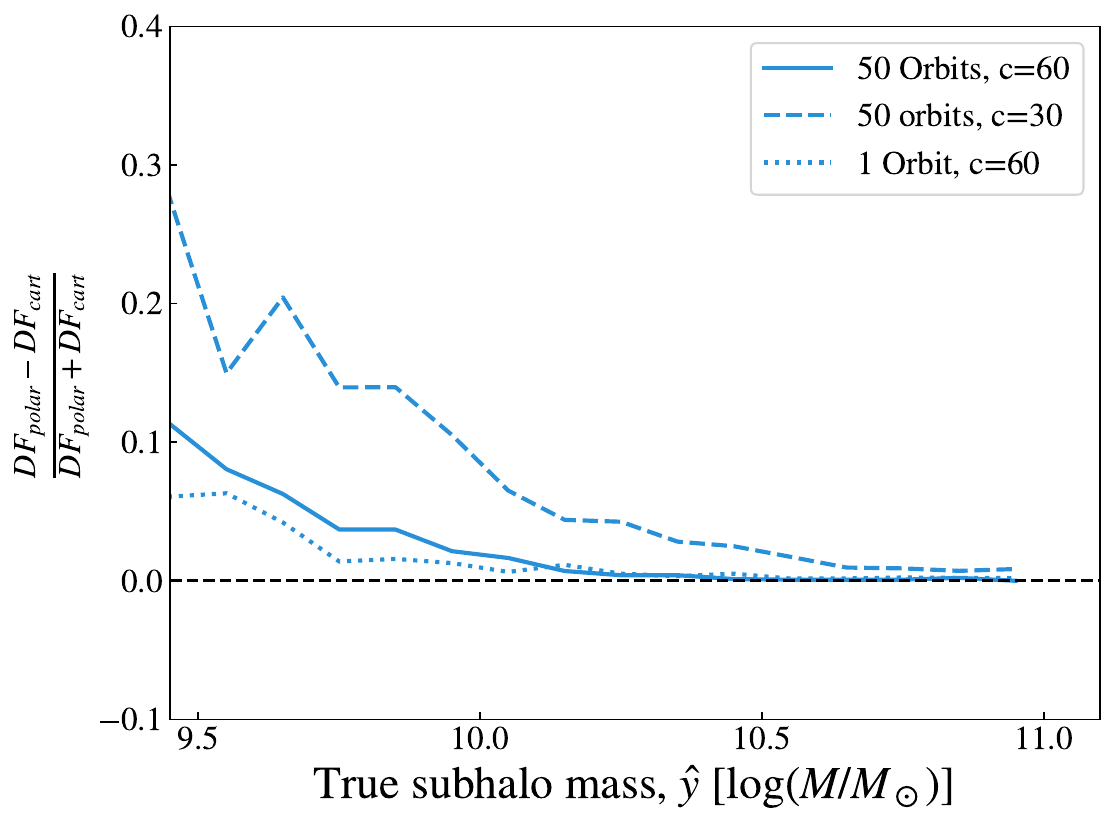}
    \caption{The relative improvement of polar transforms on network performance for fiducial data (solid line), data containing low concentration subhalos (dashed line) and data with low signal-to-noise (dotted line). Positive values indicate the polar transform improves network performance and negative values indicate it diminishes network performance. The figure only extends to masses where the statistic behaves sensibly due to extremely limited detections at low masses.}
    \label{fig:hard_regime_testing}
\end{figure}

Figure \ref{fig:hard_regime_testing} shows this statistic for the ideal data, low signal-to-noise data, and low subhalo concentration data as a function of subhalo mass. For analysis, we consider only $\log(M/M_\odot) \geq 9.5$ as, below this mass, detections are too scarce to draw conclusions. As subhalo mass decreases, the statistic for all three data types increases from 0, showing the relative improvement of detection fraction offered by polar transforms increases for smaller perturbations. Compared with the fiducial data (50 HST orbits, $c=60$), this effect is amplified for $c=30$, however slightly diminished for low SNR data.

Low concentration subhalos, for which the polar transform has the most effect, produce much more diffuse and spatially extended perturbations. Therefore, the geometric linearisation and receptive field arguments are likely much more potent, as more convolution strides are required to capture the full extent of the perturbation. The increase in relative improvement is diminished for the low SNR data, but still does increase as subhalo mass decreases. This is likely due to the decreased SNR suppressing perturbations equally for the Cartesian and polar representations. Therefore, the geometric arguments are still valid, but cannot compensate for completely lost signal from increased noise.

\subsection{Limitations and caveats}

While it has been shown that the polar transform is effective in increasing network performance over several regimes, we note here several caveats that must be considered.

\begin{itemize}
    \item We argue above that approximately circular ring morphologies in Cartesian space transform into linear morphologies in polar space. This benefits the network in two ways; reducing degenerate angular features allowing for better feature learning, and arranging the data into a format more conducive to rectangular convolutions. Therefore, in the circumstance where this approximate circularity breaks down, the shape of the ring in the polar representation may be equally as complex as in the Cartesian representation leading to diminished gains in network performance. This can arise from highly elliptical mass distributions, multi-source systems and systems with strong external shear. However, for standard galaxy-scale lenses as tested in this work, the polar transform provides an advantage.
    
    \item When we simulate images, we sample the source and lens position from priors fairly tightly bound to the centre of the image. The result is that the Einstein ring itself is also approximately central within the image. Therefore, we have thus far not needed to consider the placement of the resampling grid to perform the polar transform. If applied to real observations, the grid would have to be manually positioned at the approximate centre of the image. If done incorrectly, it could cause the ring to no longer map to a horizontal structure in polar space which could nullify the effects discussed above. The key consideration is that polar transforms are only rotationally invariant when the original image is approximately rotationally symmetric. Therefore, for the polar transform to be successfully applied, the resampling grid must be placed appropriately.
    
    \item At two stages during the training process, we use interpolation; once during the re-gridding for the polar transform, and another when using data augmentation to rotate images. In the context of subhalo detection, the relevant features are generally extremely small. Interpolation of this nature may risk smoothing the subhalo perturbation where it would otherwise be detectable. However, this does not appear to be a major factor here, as it would primarily affect polar-transformed data due to the additional interpolation, yet it maintains superior performance in all cases considered here. Another potential effect of interpolation is the introduction of correlated noise. If different grid positions have different spatial extents, the pixel values of these positions can span several pixels in polar space, leading to correlated noise. Again, there is a potential risk that this could lead the network to learn false patterns, or lead to false detections if the noise mimics a subhalo perturbation. However, we encountered no significant population of false positive predictions, suggesting this risk is minimal.
    
    \item This work includes only one subhalo per image. In this work, we show that the polar transform improves the detectability of the subhalo for convolutional networks. For approaches that characterise multiple subhalos such as image segmentation, or population level statistics such as neural ratio estimators, the advantages offered by the polar transform are unclear without further investigation. This is because the introduction of multiple subhalos changes the geometry such that perturbations can overlap, and the morphology increases in complexity. The geometric arguments we use above may not identically apply within this regime. 
\end{itemize}

\subsection{Alternative Approaches/Architectures}

In this work, we have tested the polar transform on one machine learning architecture demonstrating improvement. We have argued that the gain in performance is primarily tied to CNN biases by aligning the data to be more conducive with how convolutions extract features. It should therefore be the case that other architectures that utilise convolutions will benefit to some degree. This will apply to segmentation tasks that use U-Net architectures, as well as the feature extractors in Neural Ratio Estimators used to extract population information.

A potential alternative to polar transformations are group equivariant CNNs (G-CNNs). These networks utilise convolutional layers that respond consistently to rotated versions of the same feature, essentially natively replicating what the polar transform was designed to do.

One architecture which may not respond well to the polar transform are transformer based architectures such as Vision Transformers (ViTs). These architectures rely less on convolutions, and more on attention and patching to directly model global relationships between features. It is therefore less likely to benefit from the alignment of receptive fields that the polar transform will offer, however may still benefit from rotational invariance provided by linearising the ring.

\section{Conclusions}\label{sec:conclusions}
In this work, we explored the potential of using polar transforms to increase network performance for the task of subhalo detection within simulated galaxy scale strong lensing images. We also examined the effect of using a network with pretrained weights instead of randomly initialized weights.

We first conducted our comparison using idealized data with high SNR and subhalos with concentration $c=60$. For this we trained four networks, one for each combination of weight initialization and coordinate representation. We found that the polar transform had superior performance for both pretrained and randomly initialized networks across the full range of tested subhalo masses. Additionally, the pretrained networks consistently outperformed their randomly initialized counterparts. In the best case, the increase in detection fraction for the pretrained networks was $\sim15$ per cent, and $\sim50$ per cent for the randomly initialized networks.

We extended our analysis to more realistic data, first by reducing the SNR and using $c=60$ subhalos, then by using high SNR but reducing the subhalo concentration to $c=30$. For both of these, we trained one network on polar-transformed data and one on Cartesian data, each with pretrained weights. In the case of the low SNR data, the improvement became much smaller than with the idealised data, having an improvement in detection fraction of $\sim1-5$\,per cent, with the improvement almost exclusively occurring in systems with low source magnitudes. For the data with low concentration subhalos, the network using polar transformed data maintained a significant advantage over its Cartesian counterpart, with detection fractions increasing $10-15$\,per cent for $9 \leq \log(M/M_\odot)\leq 10$. We also found that the advantage provided by the polar transform increases as the subhalo perturbations become harder to detect, either due to increased noise or smaller perturbation. These results suggest that polar coordinate representations provide a simple and computationally inexpensive means of improving CNN-based analyses of strong lensing data, particularly in regimes where subhalo perturbations are difficult to detect.

For this work, we used a consistent architecture across all comparisons. We are therefore unable to draw conclusions on whether the increased network performance from the polar transform will generalise to other network architectures. Future work should investigate whether similar gains can be achieved for alternative architectures such as U-Net-based segmentation networks, neural ratio estimators, and transformer-based models, as well as on more realistic lensing simulations and observational data.

\section*{Acknowledgements}

EC acknowledges the support of a Royal Society Dorothy Hodgkin Fellowship Award DHF\textbackslash R\textbackslash 241008. JC acknowledges the support of the same award for the funding of his PhD studentship. SD acknowledges support by the Science and Technology Facilities Council [grant number ST/X000982/1].

\section*{Data Availability}

All data and analysis code used in this work are available from the first author on reasonable request.



\bibliographystyle{mnras}
\bibliography{example} 




\appendix


\bsp	
\label{lastpage}
\end{document}